\documentclass{PoS}

\usepackage{epsfig}
\usepackage{subfigure}

\renewcommand{\thesubfigure}{\arabic{subfigure}}
\makeatletter
   \renewcommand{\p@subfigure}{}
   \renewcommand{\@thesubfigure}{Fig. \thesubfigure:\hskip\subfiglabelskip}
\makeatother

\def\Journal#1#2#3#4{{#1} {\bf #2}, #3 (#4)}

\def\NPB{{\em Nucl. Phys.} B}
\def\PLB{{\em Phys. Lett.}  B}
\def\PRL{\em Phys. Rev. Lett.}
\def\PRD{{\em Phys. Rev.} D}

\title{Status of the global fit to electroweak precisions data}

\ShortTitle{Global Electroweak Fit and Constraints on the Higgs Mass}

\author{\speaker{Martin Goebel~\thanks{on behalf of the Gfitter Group (www.cern.ch/Gfitter).}}\\
        DESY and Institut f\"ur Experimentalphysik der Universit\"at Hamburg,\\
        E-mail: \email{martin.goebel@desy.de}}

\abstract{In this presentation Gfitter results from the global Standard Model (SM) fit to electroweak 
precision data are discussed. We have used the latest measurements of $m_{\rm{top}}$ and $M_W$ and 
the most recent results for direct Higgs searches at LEP and Tevatron. We obtain 
$M_H=121^{\;+\,17}_{\;-\,6}\:\mathrm{GeV}$ and a 95\% CL upper limit of 155 GeV for the SM Higgs mass. 
The forth-order result for the strong coupling constant is given by  
$\alpha_S(M_Z^2)=0.1193 \pm 0.0028 {\rm(exp)} \pm 0.0001{\rm (theo)}$. 
In addition the electroweak fit has been performed with the 
top mass determined from the $\ensuremath{p\overline p}\to\ensuremath{t\overline t}+X$ 
cross-section as measured at Tevatron.}

\FullConference{35th International Conference of High Energy Physics - ICHEP2010,\\
		July 22-28, 2010\\
		Paris France}

\begin{document}

\section{Introduction}
Precision measurements allow us to probe physics at much higher energy scales than 
the masses of the particles directly involved in experimental reactions by 
exploiting contributions from quantum loops. Prominent examples are the electroweak 
precision measurements, which are used in conjunction with the Standard Model (SM)
to predict via multidimensional parameter fits unmeasured quantities like the Higgs mass.
Such an approach has been used in the {\em Gfitter} analysis of the Standard Model (SM) in
light of electroweak precision data~\cite{Gfitter}.\par
In this paper updated results of the global electroweak fit are presented
taking into account the latest experimental precision measurements and the results of
direct Higgs searches from LEP and Tevatron.
\section{Fit Inputs}
The SM predictions for the electroweak precision observables measured by the LEP, SLC, 
and Tevatron experiments are fully implemented in {\em Gfitter}. State-of-the-art calculations 
are used, in particular the full two-loop and leading beyond-two-loop corrections 
for the prediction of the $W$ mass and the effective weak mixing 
angle~\cite{awramik}, which exhibit the strongest 
constraints on the Higgs mass. The Gfitter library also includes the fourth-order
(3NLO) perturbative calculation of the mass-less QCD Adler function~\cite{adler},
allowing the fit to determine the strong coupling constant with negligible theoretical 
uncertainty.\par
The experimental data used in the fit include the electroweak precision data 
measured at the $Z$ pole including their experimental correlations~\cite{zsummary},
the latest $W$ mass world average $M_W=(80.399\pm0.023)\:\mathrm{GeV}$~\cite{wmass}  
and width $\Gamma_W=(2.098\pm0.048)\:\mathrm{GeV}$~\cite{wwidth}, and the newest average of
the Tevatron top mass measurements $m_{\rm{top}}=(173.1\pm1.3)\:\mathrm{GeV}$~\cite{topmass}. For the
contribution of the five lightest quark flavours to the electromagnetic coupling 
strength at $M_Z$ we use the evaluation from~\cite{hagiwara}.  In addition, for some results we take 
also into account the information from the direct Higgs searches at LEP~\cite{lepHiggs} 
and Tevatron~\cite{tevHiggs}.

\begin{figure}[h]
\centering
\subfigure[$\Delta \chi^2$ as a function of $M_H$ for a fit without the direct Higgs searches from 
LEP and Tevatron. The red solid line shows the result for a fit using the top mass as determined from 
the $\ensuremath{t\overline t}$ cross-section.]
   {\epsfig{figure=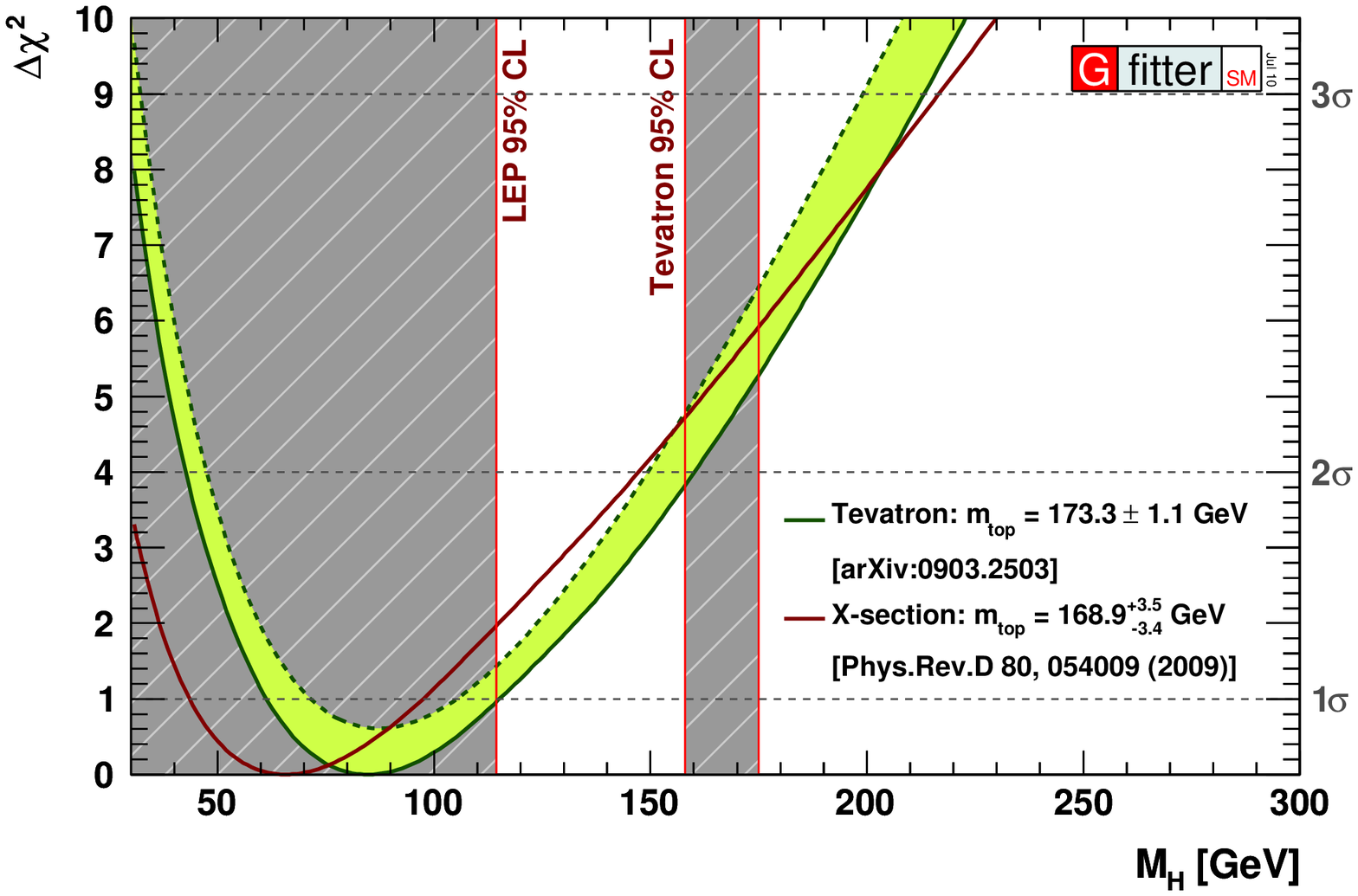, scale=0.35}\label{fig:Higgs1}}\hfill
\subfigure[$\Delta \chi^2$ as a function of $M_H$ for a fit including the direct Higgs search results from 
LEP and Tevatron.]
   {\epsfig{figure=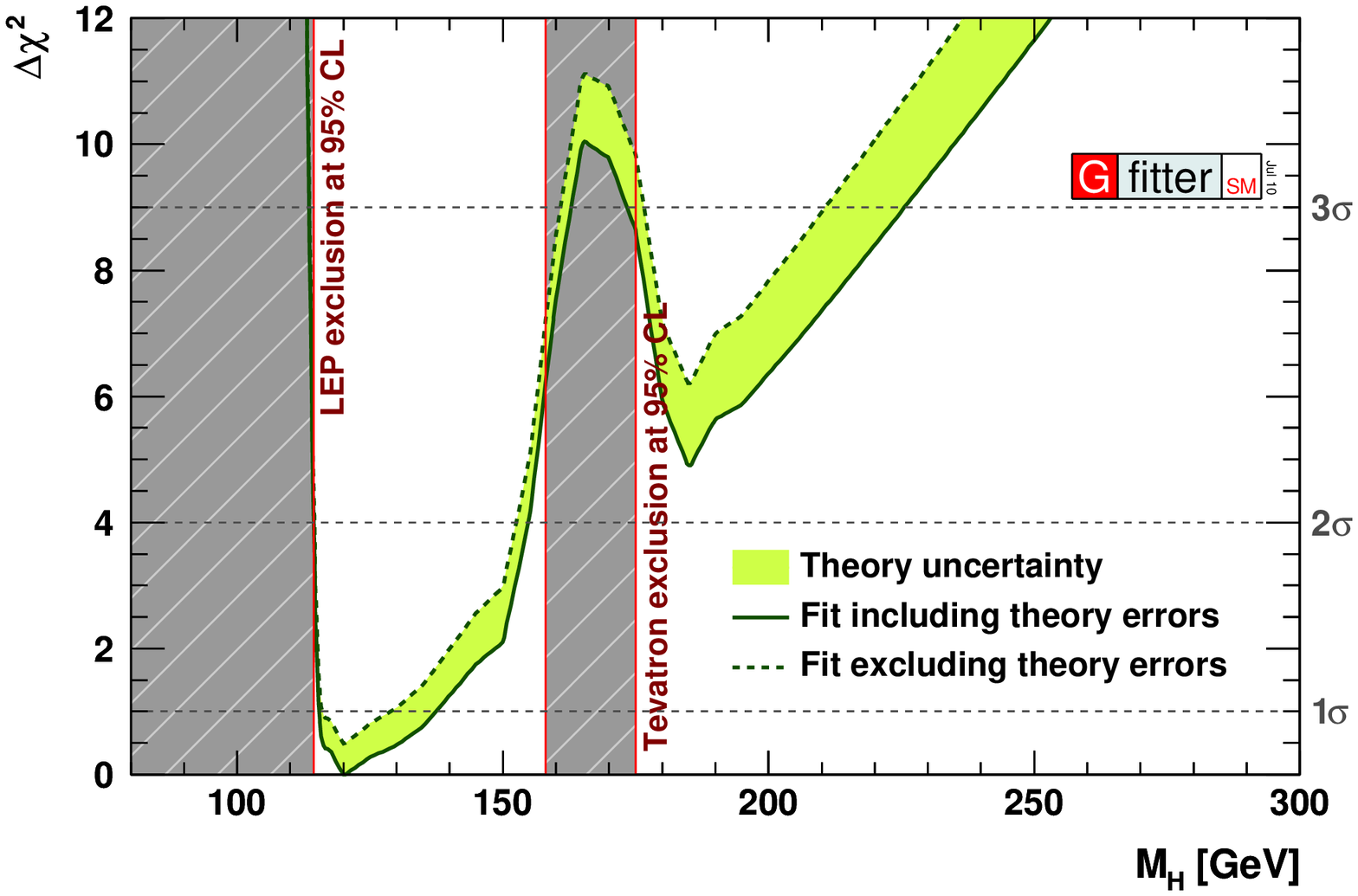, scale=0.35}\label{fig:Higgs2}}
\end{figure}

\section{Fit Results}
The minimum $\chi^2$ value of the fit with (without) using the information of the direct 
Higgs searches amounts to 17.8 (16.4) which corresponds to a {\it p}-value of 0.23 (0.22).    
Figure~\ref{fig:Higgs1} and fig.~\ref{fig:Higgs2} show the corresponding profile curves of the 
$\Delta \chi^2$ estimator. We find for the most probable Higgs mass the value 
$M_H=84^{\;+\,30}_{\;-\,23}\:\mathrm{GeV}$ ($M_H=121^{\;+\,17}_{\;-\,6}\:\mathrm{GeV}$). 
The 95\%  upper limits are 159 GeV (155 GeV).\par
\begin{figure}[h]
\centering
\subfigure[Contours of 68\%, 95\% and 99\% CL obtained from scans of fits with fixed variable pairs of
$m_{\rm{top}}$ and $M_H$.]
   {\epsfig{figure=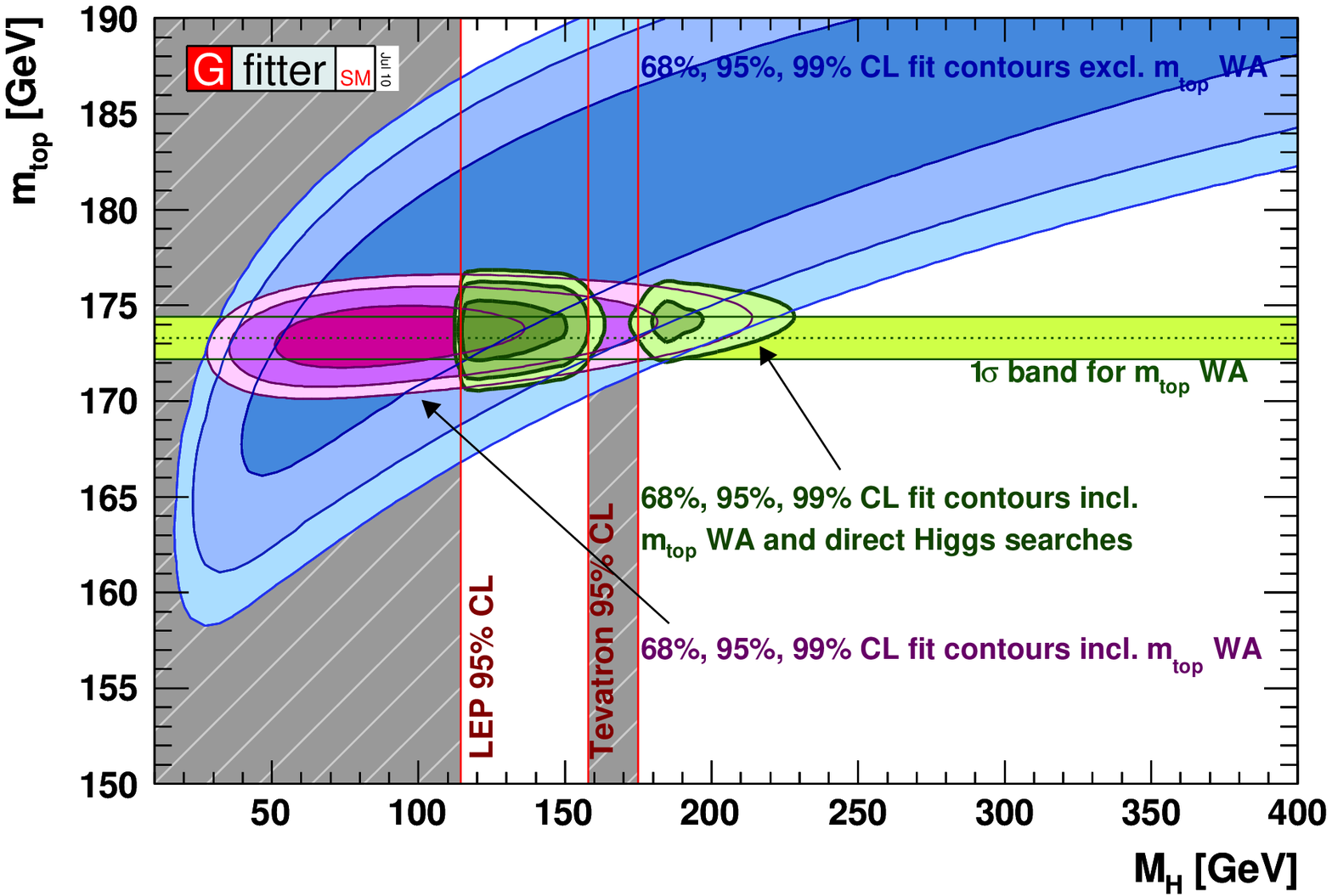, scale=0.35}\label{fig:TopHiggs}}\hfill
\subfigure[$\Delta \chi^2$ as a function of $m_{\rm{top}}$ for a fit with and without the direct Higgs searches.]
   {\epsfig{figure=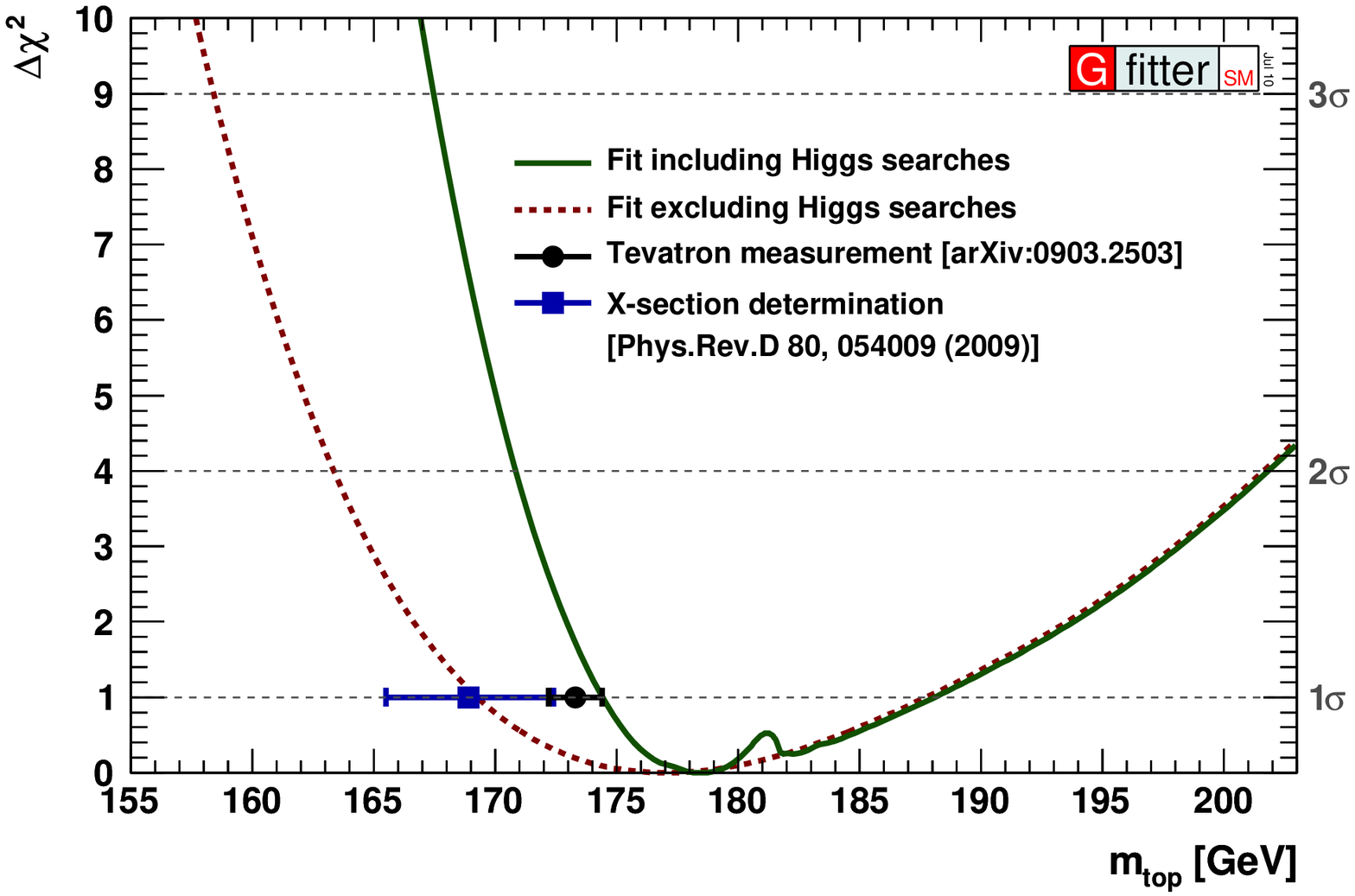, scale=0.35}\label{fig:Top}}
\end{figure}
Figure~\ref{fig:TopHiggs} shows the 68\%, 95\% and 99\% CL contours for the variable pairs of 
$m_{\rm{top}}$ vs. $M_H$. Three sets of fits are shown: the largest/blue (narrower/purple) allowed 
regions are derived from a fit excluding (including) the measured top mass value 
(indicated by the shaded/light green horizontal band). The fit providing the narrowest constraint (green)
uses all available information, {\it ie}., including the direct Higgs searches from LEP and Tevatron. 
The largest/blue contour show nicely the positive correlation factor between the Higgs and top mass, 
which can be determined to be 0.31. The importance of the top mass for the Higgs mass determination 
is clearly visible.
However, an additional uncertainty for the top mass could arise due to ambiguities in the top mass 
definition at Tevatron~\cite{Hoang}. In an alternative approach the top mass has been determined from the SM 
$\ensuremath{p\overline p}\to\ensuremath{t\overline t}+X$ cross-section~\cite{xsection}, where 
the top mass is unambiguous once a renormalisation scheme is defined.
The use of this top mass in the electroweak fit leads to a smaller value of $M_H$, but due the larger error 
of this top mass determination the 95\% and 99\% CL upper limits does not change significantly 
(see fig.~\ref{fig:Higgs1}). Figure~\ref{fig:Top} shows the $\Delta \chi^2$ as a function of $m_{\rm{top}}$. For a comparison the direct Tevatron measurement and the determination from the $\ensuremath{t\overline t}$
cross-section are also shown.\par
\begin{figure}[t]
\centering
\subfigure[Determination of $M_H$ excluding all other sensitive observables from the fit, except for the one given.]
   {\epsfig{figure=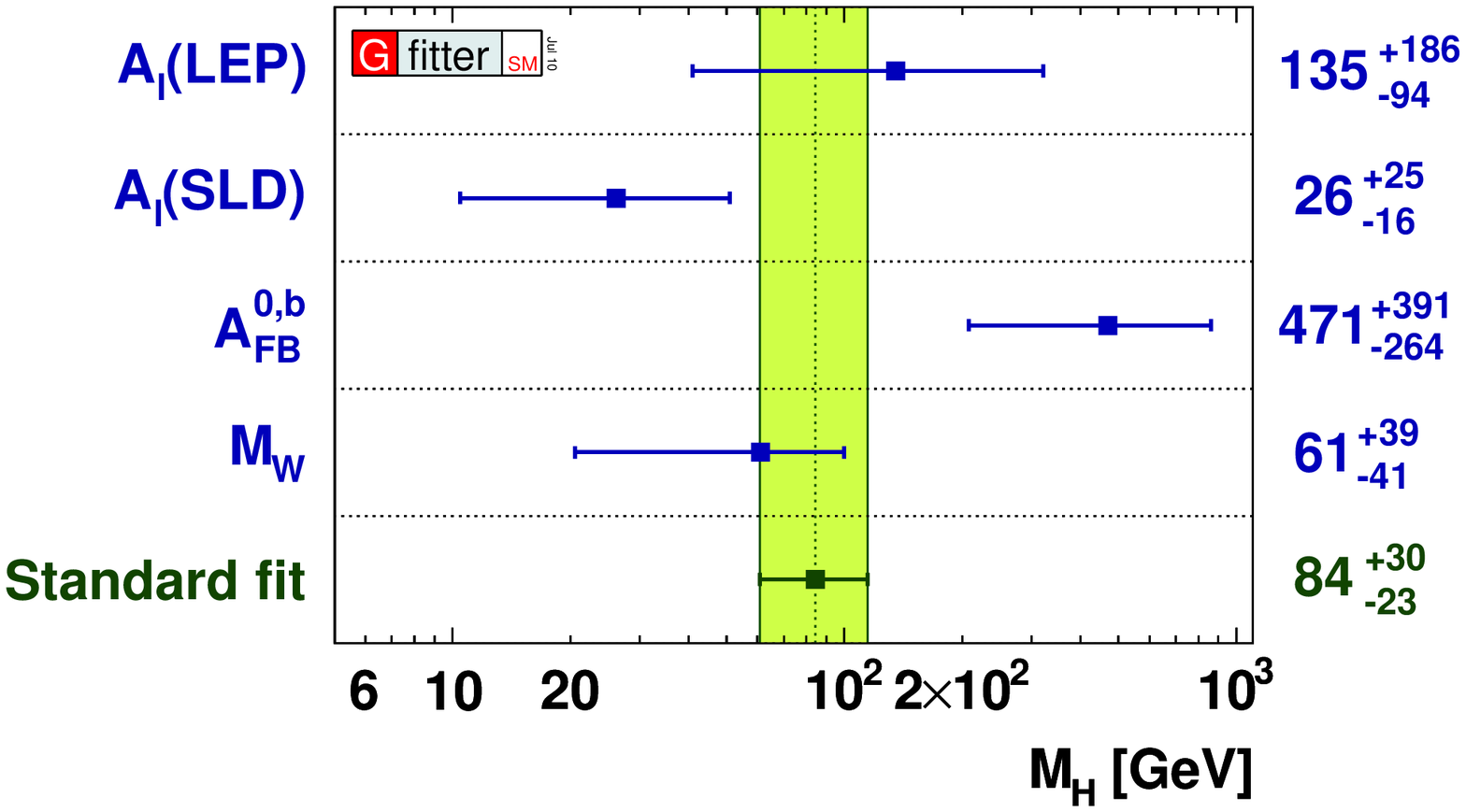, scale=0.37}\label{fig:MainObs}}\hfill
\subfigure[Top: Collection of  $\alpha_s(\mu)$ measurements at order 3NLO, 2NLO, and NLO~\cite{alphaQCD}. 
Bottom: The corresponding 
         $\alpha_s$ values evolved to $M_Z$~\cite{alphaQCD}.]
   {\epsfig{figure=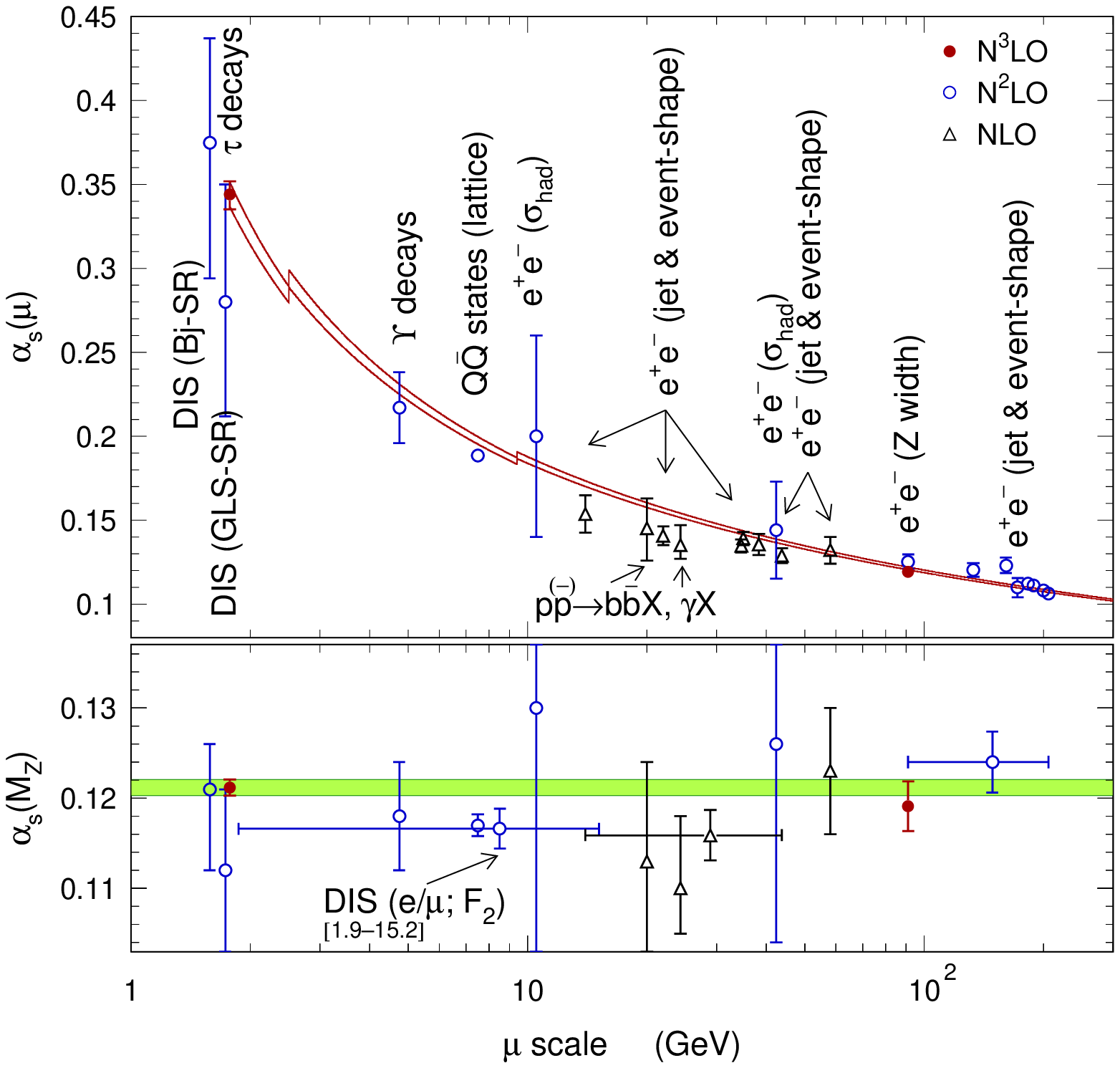, scale=0.34}\label{fig:QCD}}
\end{figure}

In fig.~\ref{fig:MainObs} only the observable indicated in a given row is included in the fit.
The four observables providing the strongest constraint on $M_H$ are shown. 
The compatibility among these measurements can be estimated
by repeating the global fit where the least compatible of the measurements (here
$A_{\rm FB}^{0,b}$) is removed, and by comparing the $\chi^2_{\rm min}$ estimator obtained in that fit to the
one of the full fit. To assign a probability to the observation, the
$\Delta\chi^2_{\rm min}$ obtained this way must be gauged with toy MC experiments to take into account the
``look-elsewhere'' effect introduced by the explicit selection of the outlier. We find that
in $(1.4\pm0.1)\%$ (``$2.5\sigma$'') of the toy experiments, the $\Delta\chi^2_{\rm min}$ exceeds the
value observed in the current data.\par
From the fit including the direct Higgs searches we find for the strong coupling at the $Z$-mass scale
$\alpha_s(M_Z^2)=0.1193^{\,+0.0028}_{\,-0.0027} \pm 0.0001$, where the first error is experimental 
and the second due to the truncation of the perturbative QCD series. Figure~\ref{fig:QCD} shows the 
excellent agreement between our result and 3NLO result from $\tau$ 
decays~\cite{alphaQCD}.


\begin{thebibliography}{99}
\bibitem{Gfitter}
H.~Fl\"acher {\it et.al.} {\it Eur.~Phys.~J.} C {\bf 60} (2009) 543, [arXiv:0811.0009 [hep-ph]].

\bibitem{awramik}
M.~Awramik {\it et al.}, \Journal{\PRD}{69}{053006}{2004}, [hep-ph/0311148];
M.~Awramik {\it et al.}, {\it JHEP} {\bf 11}, 048 (2006), [hep-ph/0608099];
M.~Awramik {\it et al.}, \Journal{\NPB}{813}{174}{2009}, 048 (2006), [arXiv:0811.1364 [hep-ph]].

\bibitem{adler}
P.~A. Baikov {\it et al.} \Journal{\PRL}{101}{012002}{2008}, [arXiv:0801.1821 [hep-ph]]

\bibitem{zsummary}
LEP and SLD Elektroweak and Heavy Flavour Working Groups, Phys. Rept. {\bf 427}, 257 (2006), [hep-ex/0509008].
\bibitem{wmass}
Tevatron Electroweak Working Group and CDF and D0 Collaboration, [arXiv:0908.1374 [hep-ex]].

\bibitem{wwidth}
Tevatron Electroweak Working Group Collaboration, [arXiv:1003.2826 [hep-ex]].

\bibitem{topmass} 
CDF and D0 Collaboration, arXiv:1007.3178 [hep-ex]], Elizaveta Shabalina, these proccedings.

\bibitem{hagiwara} 
K.~Hagiwara {\it et al.}, \Journal{\PLB}{649}{173}{2007}, [hep-ph/0611102].

\bibitem{lepHiggs}
LEP Working Group for Higgs boson searches, Phys.\ Lett.\  {\bf B565 } (2003)  61-75., [hep-ex/0306033].

\bibitem{tevHiggs}
CDF and D0 Collaboration, [arXiv:1007.4587 [hep-ex]], Ben Kilminster, these proceedings.

\bibitem{Hoang}
  A.~H.~Hoang, I.~W.~Stewart,
  Nucl.\ Phys.\ Proc.\ Suppl.\  {\bf 185 } (2008)  220-226.
  [arXiv:0808.0222 [hep-ph]].

\bibitem{xsection}
  U.~Langenfeld, S.~Moch, P.~Uwer,
  Phys.\ Rev.\  {\bf D80 } (2009)  054009.
  [arXiv:0906.5273 [hep-ph]], Peter Uwer, these proceedings

\bibitem{alphaQCD}
  M.~Davier, S.~Descotes-Genon, A.~Hocker, B.~Malaescu and Z.~Zhang,
  Eur.\ Phys.\ J.\  C {\bf 56} (2008) 305
  [arXiv:0803.0979 [hep-ph]].
  
\end{thebibliography}
\end{document}